# Fuzzy Logic-Based System for Brain Tumour Detection and Classification

NVSL Narasimham[1] ,PhD., Keshav Kumar K.[2],M.Sc.

[1]Department of Humanities and Mathematics,G. Narayanamma Institute of Technology and Science (for Women), Hyderabad-500104, India

[2]Department of Humanities and Mathematics, G. Narayanamma Institute of Technology and Science (for Women), Hyderabad-500104, India

*Corresponding Author's Email: keshav.maths@gnits.ac.in

***Abstract***

Brain Tumours (BT) are extremely dangerous and difficult to treat. Currently, doctors must manually examine images and manually mark out tumour regions to diagnose BT; this process is time-consuming and error-prone. In recent times, experts have proposed automating approaches for detecting BT at an early stage. The poor accuracy and highly incorrect prediction results of these methods caused them to start the research. In this study, we suggest a fuzzy logic-based system for categorising BT. This study used a dataset of 253 Magnetic Resonance Imaging (MRI) brain images that included tumour and healthy images. The images were first pre-processed. After that, we pull out features like tumour size and the image's global threshold value. The watershed and region-growing approach is used to calculate the tumour size. After that, the fuzzy system receives the two features as input. Accuracy, F1-score, precision, and recall are used to assess the results of the fuzzy by employing both size determination approaches. With the size input variable discovered by the region growth method and global threshold values, the fuzzy system outperforms the watershed method. The significance of this research lies in its potential to revolutionize brain tumour diagnosis by offering a more accurate and efficient automated classification system. By reducing human intervention and providing reliable results, this approach could assist medical professionals in making timely and precise decisions, leading to improved patient outcomes and potentially saving lives. The advancement of such automated techniques has the potential to pave the way for enhanced medical imaging analysis and, ultimately, better management of brain tumour cases.

## INTRODUCTION

BT forms from abnormal groupings of cells in brain tissue. BT is classified as benign or malignant. Malignant BT is among the deadliest types of cancer and can be lethal even in the absence of other symptoms [1], whereas benign BT can be surgically treated. Primary BT begins in another part of the body and spreads to the brain, whereas metastatic BT begins in the brain. The most common BT in adults is lymphoma and gliomas, with gliomas accounting for more than 80% of malignant BT and developing from periglial tissue. The location of the lesion causes headaches, nausea, impaired eyesight, convulsions, and mental fog. There is a more sophisticated classification system for BT, which separates them into four groups, with higher grades reflecting more aggressive tumours. According to the 2020 Global Cancer Statistics [2], there will be around 308,000 new diagnoses of BT in 2020 and about 251,000 deaths from BT.

Early identification of BT is critical for effective therapy. As medical imaging has evolved, imaging techniques have come to play a vital role in the diagnosis and evaluation of BT treatment. BT can be difficult to identify, but new imaging tools can help doctors understand more about the tumour's size, location, and shape. MRI is a common scanning technique in neurology. MRI employs radiofrequency impulses to stimulate the tissue while it is subjected to an extremely strong magnetic field to obtain an image of its inside. Soft tissue contrast is improved while no radiation is needed, making it a win-win situation. As a result, MRI is superior for diagnosing brain malignancies (Bunevicius et al., 2020).

The recent triumphs of AI in intelligent medicine have brought the topic to the forefront. Recently, research in medical image processing has focused on employing AI approaches to categorise and segment MRI. BT classification has two basic applications: evaluating whether an image of the brain is normal or abnormal depending on the presence or absence of a tumour, and then classifying within aberrant brain images to distinguish between various types of tumours.

The goal of the research is to create an efficient and automatic system for classifying MRIs of BT to assist doctors in their diagnostic processes. The key benefits of the provided method are listed below.

- Kaggle data and photos are collected and sent into the pre-processing step.
- Watershed and region growth are two ways of determining the boundary.
- Using Fuzzy Logic, an image can be classified as "normal" or "abnormal".
- The efficiency of two-edge detection with fuzzy classification algorithms is assessed using accuracy, recall, precision, and F-score.

The remaining paper is structured as follows: In Section 2, we will examine existing research on possible techniques for classifying BT. The proposed methodology is laid out in detail in Section 3. Data and processing methods are described in further depth in Section 4. The operation of a fuzzy logic system is discussed in Section 5. The experiment results are discussed in Section 6. Section 7 contains the conclusion and analysis.

## LITERATURE SURVEY

A significant number of the researchers have previously worked on AI-based BT detection. Some of their work is detailed further below.

Shanthi S. et al.(2022) proposed an automatically optimised hybrid deep neural network (OHDNN) for the diagnosis of BT. The suggested method's two key components are pre-processing and BT categorization. After being built from raw data, the images go through a series of pre-processing phases. After being pre-processed, the pictures are submitted to a categorization stage. For the classification method in this study, OHDNN is used. The HDNN is built on a convolutional neural network with long short-term memory (CNN-LSTM). According to the testing results, the proposed approach has 97.5 percent accuracy. Kang et al.(2021)

suggested a method for recognising and diagnosing BT by combining deep characteristics with Machine Learning (ML) classifiers. They present a method for extracting these hidden features from brain MRI scans using Transfer Learning (TL) and a large number of pre-trained deep Convolutional Neural Networks (CNN). Multiple ML classification methods are used to assess the extracted deep features. Specifically for large datasets, studies shows that support vector machine (SVM) perform better than regular ML classifiers. Rasool et al. (2022) used MRI data to construct a unique hybrid CNN for BT classification. In this research, they offer a technique that combines Deep Learning (DL) and CNN to achieve better classification results. The first approach integrates a pre-trained Google Net of the CNN algorithm with the SVM categorization technique. The second technique pairs a SoftMax classifier with a finely tuned Google Net. The proposed method was tested on brain MRIs containing three types of tumours. Enhanced recognition accuracy was achieved by employing Google-Net as a feature extractor and an SVM model.

Gu, Xiaoqing et al. (2021) proposed that magnetic resonance images of BT be classified using convolutional dictionary learning with local constraint (CDLLC). To probe discriminative data, they use CNN architecture with multi-layer dictionary learning. Encoding a vector on a dictionary yields a sparse coding vector, which can be thought of as multiple projections onto new spaces. Meanwhile, they produce a local constraint of atoms using a supervised k-nearest neighbour network, utilising the supervised information to ensure that the created dictionary has strong discriminating. To address the presented issue, an effective iterative optimization technique is proposed. Using the Cheng and Rembrandt datasets, the experiment creates two therapeutically relevant multi-class classification approaches. The results of the evaluation suggest that the suggested approach is effective for classifying MRI of BT and has the potential to outperform other comparisons. In the study [9], they created a powerful method for classifying BT using DL approaches as a response to the inaccuracy of current artificial diagnosis systems (Haq et al., 2022). The proposed method classifies MRI data using an augmented CNN. Data augmentation and TL were employed to enhance the model's classification performance. The results validated the model's great accuracy when compared to the reference models. Because of its good predictive findings, they recommend the suggested model for BT diagnosis in IoT-healthcare systems.

BT is often considered to be the deadliest and most destructive illness due to its singular combination of features, poor survival rate, and aggressive nature. Challenges in identifying and treating BT may be solved through DL. Solving the issue of centralised data collection is the focus of this effort. The goals of this research were to develop a VGG-16 based CNN framework for the detection of BT and to find the training parameters. Their approach may also work with MRI to spot BT. The results of the evaluation showed that the approach performed better than the established methods of diagnosis for BT. Zain Eldin et al. (2022) purposed an improvised model so that it may be used to better diagnose BT. Present a CNN constructed on top of a BT Classification Model (BCM-CNN) in this research. To fine-tune the CNN's hyperparameters, the Adaptive Dynamic Sine-Cosine Fitness Grey Wolf Optimizer (ADSCFGWO) algorithm was utilised. Results from the experiments using the BCM-CNN as a classifier reveal that its performance was significantly improved by tuning. Incredibly high accuracy was achieved by the BCM-CNN when it was fed the BRaTS dataset. F.S Chowdhury et al. (2023) presented a strategy based on artificial neural networks (ANNs) to appropriately identify BT. To ensure homogeneity, the photographs are sized first. The watershed approach is used to segment the images once they have been normalized. The derived seven statistical features are then supplied to the ANN classifier, which is then used to categorise BT. The experimental results of the proposed method provide precision that exceeds current state-of-the-art procedures. Furthermore, the seven statistical indicators chosen are relatively limiting in presenting this performance, especially in comparison to other recent approaches.

**MATERIALS AND METHODS**

The MRI data are collected from Kaggle and processed by the median filter to remove the noise. The research flow charts for the categorisation of BT using fuzzy logic are shown in Figure 1. Next, the filtered images are processed separately to get features like global image thresholding and size. The identified features are given as input variables to the fuzzy system, and the output will be the tumour and non-tumour.

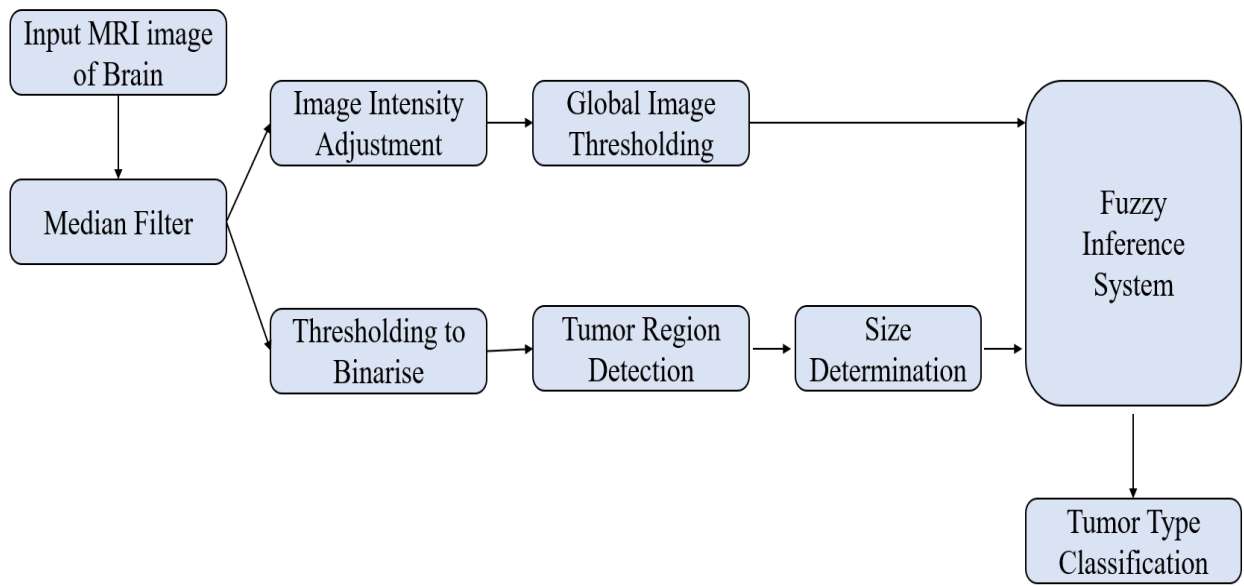

Fig.1. Research flow chart of BT classification using fuzzy system

**DATA AND ITS PROCESSING**

The MRI brain scan is pulled from the Kaggle database [13] and examined for tumour detection. The collected MRI contains 98 images of non-tumour and 155 images of the tumour. The distribution of BT data in the training and testing phases is detailed in Table 1 and Figure 2.

**Table 1.** Brain Tumour data

| MRI | Train | Test |
|---|---|---|
| Non-Tumour | 78 | 20 |
| Tumour | 124 | 31 |

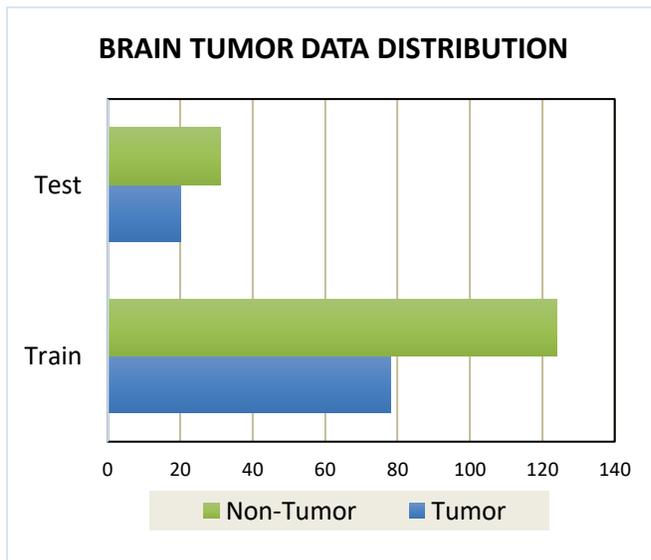

**Fig 2**. Brain Tumour data distribution

The MRIs are resized suitably in a matrix to maintain their aspect ratios. A median filter [14] is then applied to the scaled image to reduce noise while preserving the image's edges. From the filtered images the size of the BT and global threshold values are extracted.

*A. Tumour Region Detection*

The filtered image is then transformed into a binary one using a threshold. The model uses a 0.65 threshold. Next, the tumour region of the disc-shaped structure is constructed from the segmented area of the watershed and growing region approach. To locate the cancerous area, the structural element is enlarged and reassembled morphologically beneath the image mask.

In the realm of image segmentation, watershed techniques employ mathematical morphology. Each local minimum in the input MRI correlates to a distinct region in the segmentation outcome, and the region contour is obtained at the segmentation end, as stated and specified in the method of watershed approach [15]. The total count of "regions" is proportional to the count of "local minima" that exist. Since there will be more local minima in the MRI than the target item of practical value, a large number of incorrect contours is generated, making it more difficult to identify the true contour of interest. This is because of noise and local irregularities in segmented MRI. A term for this is "over-segmentation." To find the least region of the gradient image, the forced minimum calibration method is used, and then the markers recovered from that region are used to get the least value of the gradient image. The watershed approach is used to finish the image segmentation by employing the updated gradient image. Internal and external markers are used in marker-based watershed segmentation (Zheng C.X. et al.,2018). A marker acts as a connecting element. Both the internal and exterior markers are linked to their respective backgrounds. The tagging process includes both pre-processing and developing a set of criteria for selecting tags. Grey value, connectivity, size, form, and texture are just a few of the characteristics that can be employed as marker selection criteria. When the internal marks are obtained, they are employed only as the least value regions for segmentation, and the outer mark is the watershed of the segmentation outcome. Each segmented region is then subjected to additional segmentation techniques, to separate the foreground from the background.

The underlying assumption of region-growing algorithms is that adjacent pixels within a region have similar values. The standard technique is to compare how one pixel compares to its neighbours. If the cluster's similarity criteria are met, the pixel can be assigned as a neighbour to the cluster (Shrivastava et al., 2020). Noise affects all outcomes, and the importance of the similarity criterion employed cannot be overemphasised. This procedure necessitates the use of both the image and a seed set. Each segmentation target is identified by the seeds. The regions are compared to all of the nearby, unallocated pixels, and the regions grow in size over time. The departure of a pixel's intensity value from the average of that region is used to calculate similarity. To represent that area, the pixel with the smallest difference is picked. Each pixel is assigned to a certain location in a continuous process. Seeds are utilized as an additional input in the seeded region's growth; hence, segmentation results are sensitive to seed selection, and image noise can result in misplaced seeds.

The watershed and growing region edges of the tumour or cancer area are utilised to determine the initial shape and border. The method considers basic geometric forms such as circles, ovals, and partial rounds. The borders themselves can be both regular and irregular. The size is then assigned a membership value based on its shape and border (random, small, medium, etc.).

B. *Thresholding*

The median-filtered image is modified so that just 1% of the data is used at the brightest and darkest points. The updated image now has better contrast. The intensity-modified image's global threshold is calculated using Otsu's method (Schuh C and Bohm S., 2018). This approach selects the optimal cut-off, which decreases the intra-class variation of black and white pixels. The weighted sum of the variances of black and white pixels is

$$w^2(t) = \Phi(t)\sigma_1^2 + \Psi(t)\sigma_2^2 \qquad [1]$$

if we define the probability of each as $\Phi(t)$ and $\Psi(t)$, and the variances of each as $\sigma_1^2$ and $\sigma_2^2$.

When there are two classes, minimizing the intra-class variance is the same as increasing the inter-class variance. Therefore, the two classes' minimum intra-class variance expression is:

$$w^2(t^*) = w^2 - w^2(t) \qquad [2]$$

where $t^*$ is the global best threshold and $w^2$ is the combined class variances. This global threshold can be calculated with the following MATLAB command:

$$G = graythresh(IA) \qquad [3]$$

where $IA$ is the image with its intensity adjusted, and $G$ is the threshold of the global image.

### FUZZY LOGIC CONTROL

Fuzzy logic transforms human ideas into a set of mathematical principles that can be used for decision-making and problem-solving [20]. In contrast to the exact, mathematical nature of computer logic, the rules and variables used by people in making decisions are verbal and so fuzzy. These metaphorical expressions are formally expressed as Membership Functions (MF). Fuzzy logic makes choices using symbolic verbal terms rather than numerical values, relying on amorphous verbal knowledge comparable to that of humans. Fuzzy logic can be used to model the behaviour of a system that can be modelled by rules or that requires exceedingly complex nonlinear processes. Mamdani's technique is the most widely used fuzzy inference system. Fuzzification, inference, and defuzzification are the three key steps of fuzzy rule-based systems. After the crisp stage's input and output variables have been determined and mapped to linguistic variables, the fuzzification stage can begin. When the input and output variables, as well as their associated MFs, are identified and an IF-THEN rule set is formed, the fuzzy inference technique provides a fuzzy output set. After that, defuzzification is used to convert the forecast into a sharp output result.

The $G$'s threshold value and the membership value of the size for cancer or tumour regions are given into a Mamdani Fuzzy Inference System. The inputs are then subjected to fuzzification, MFs, fuzzy rules, and defuzzification. Figure 3 depicts the designed FIS as well as its technical specifications. It employs a centroid defuzzification approach, using tumour size and a global threshold as inputs and producing the cancer type as an output.

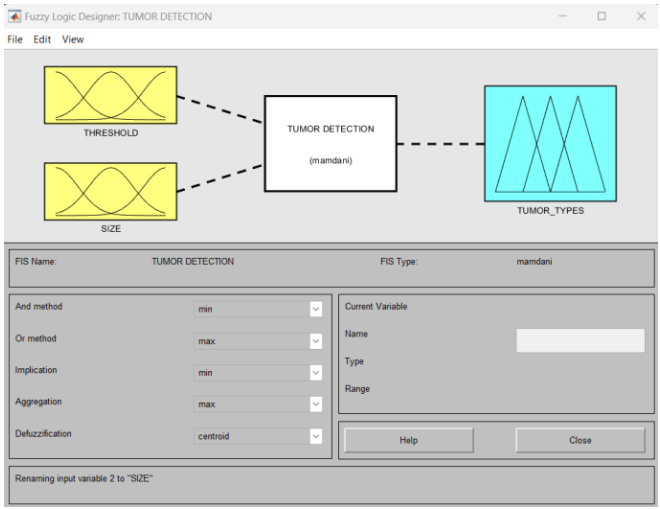

**Fig 3**. Designed fuzzy logic architecture

When the edges of BT images are examined, it is discovered that tumour cases have a round, half-round, or oval shape with an irregular border. Normal instances do not have defined boundaries. Therefore, the FIS takes this information about the shape and borders as input. After that, a membership value is established for each dimension. As a result, Figure 4 depicts the FIS MFs for the input variable "Size of BT."

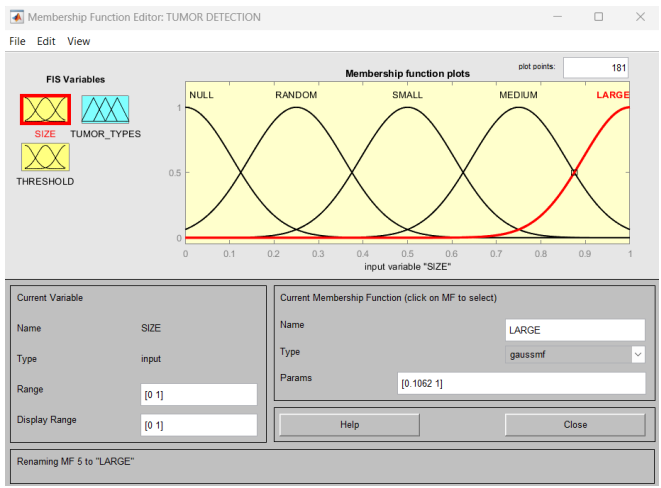

**Fig 4**. Fuzzy input variable – size

After calculating the $G$'s Threshold values of BT images, it was found that tumour instances had random, medium, and large thresholds on a scale of 1. Figure 5 depicts the MFs for the input variable "Global threshold" in the FIS, which are now assigned to each threshold scenario.

The range of the output membership value for BTs is 0.00 to 0.70, while the range for normal cases is 0.3 to 1.00. Figure 6 depicts MFs for the "Type of Tumour" output variable.

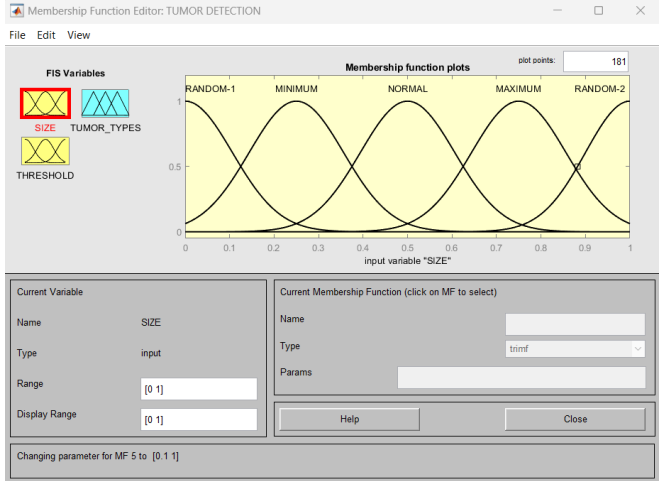

**Fig 5.** Fuzzy input variable – Threshold value

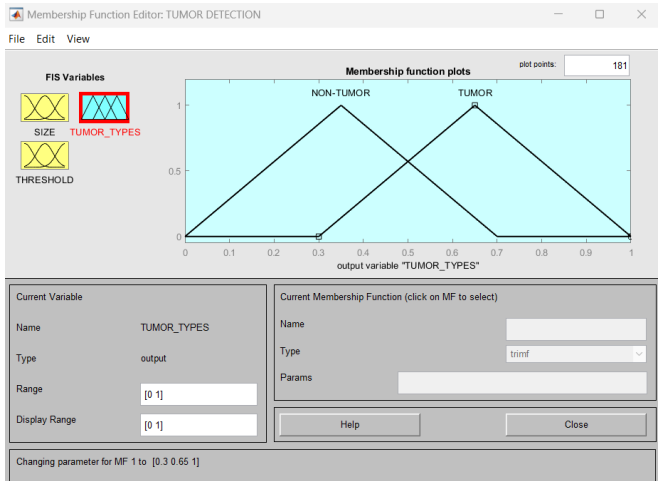

**Fig 6.** Fuzzy output variable – Tumor

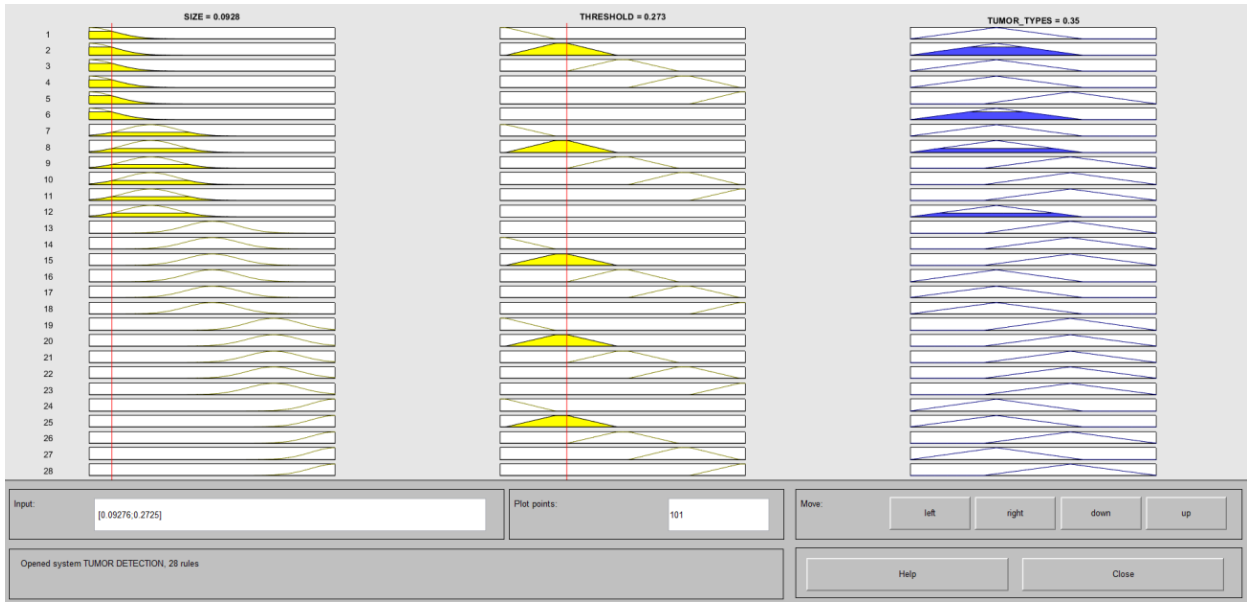

**Fig 7**. Working of fuzzy logic based on IF-THEN rules

The fuzzy working based on the IF-THEN rules is illustrated in Figure 7. Here the input variables of size and threshold values are manually changed and make the fuzzy to predict. The fuzzy gives the output value of 0.35. If the value is 0.35, then the given input corresponds to the tumour type.

## RESULT AND DISCUSSION

The outcome of BT detection using fuzzy logic is discussed in this section. The brain images from the MRI are gathered from Kaggle and they are given in figure 8. a. The raw images are resized to a particular height and width and are illustrated in Figure 8.b. Next, the images are passed through the filter to remove the noise, and the outcome of the median filter is shown in Figure 8.c. Then, the filtered images are given to the threshold techniques to convert them to binarized form. The binarized images are given in Figure 8.d. The tumour region segmentation by watershed technique is depicted in Figure 8.e.

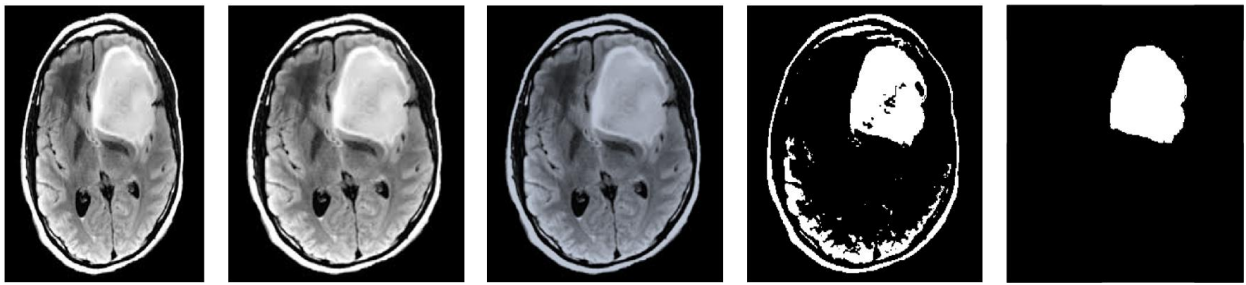

**Fig 8.** Outcome images of MRI

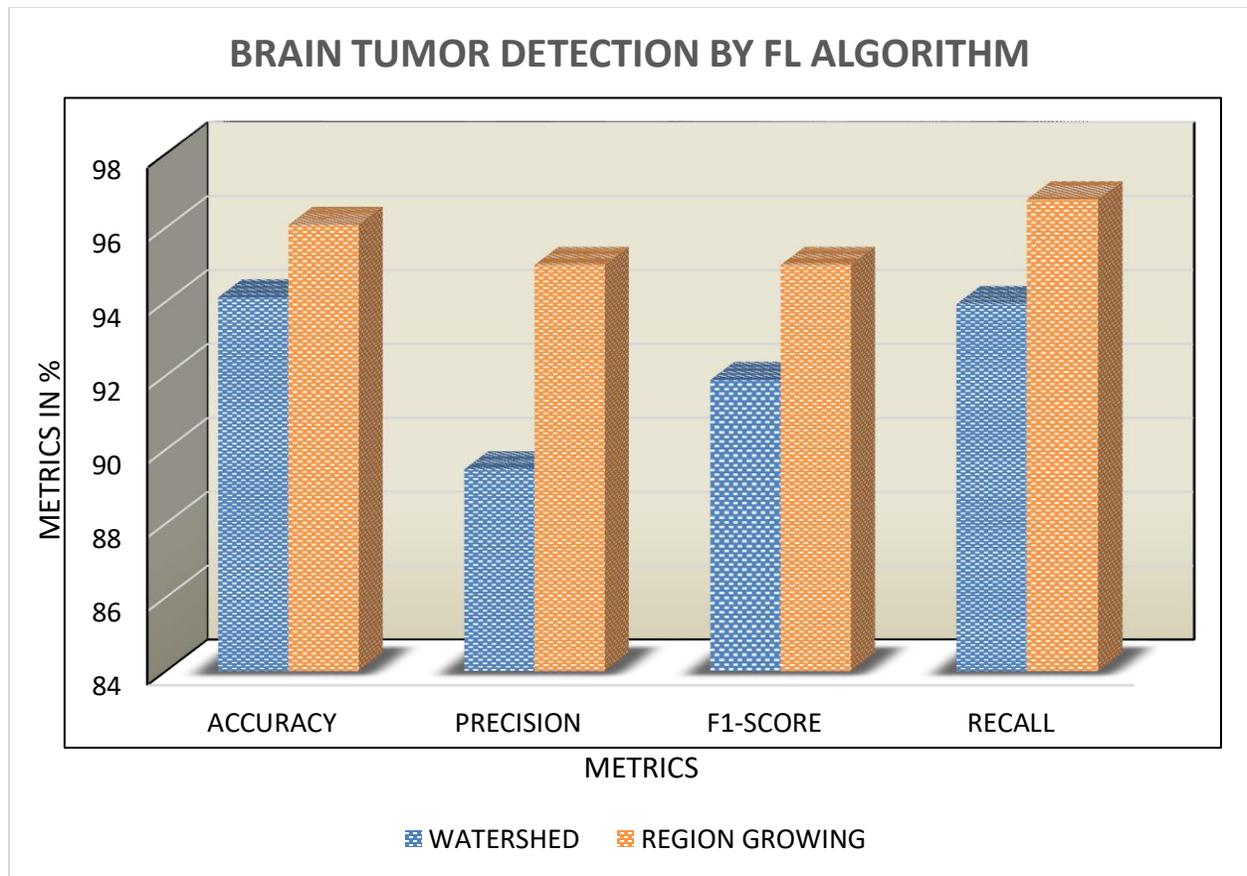

**Fig 9.** Performance evaluation of BT detection by FL Algorithm

      The fuzzy takes the input of size and threshold values as a variable. For size determination two approaches such as watershed and region growing are employed. The outcome of both methods is evaluated to find the best one. First, the size determined by the watershed is taken and given to the fuzzy logic along with the threshold values for tumour detection. For evaluation, the test data of 20 non-tumour and 31 tumour images is taken. The accuracy of this method is 94.12%, recall is 93.94%, precision is 89.47% and F1-score will be 91.89%. Next, the size determined by the growing region is taken and given to the fuzzy logic along with the threshold values for tumour detection. The same test data is used for evaluation. The accuracy of the region growing technique is 96.08%, recall is 96.77%, and precision is 95%, and F1-score will be 95%. The metrics values of both techniques are given in Table 2. Figure 8 illustrates the performance of the fuzzy system using size determined by watershed and region growing is depicted in Figure 9.

**Table 2.** Performance evaluation of BT

| Edge Detection Technique | Watershed | Region Growing |
|---|---|---|
| Accuracy | 94.12 | 96.08 |
| Precision | 89.47 | 95 |
| F1-Score | 91.89 | 95 |
| Recall | 93.94 | 96.77 |

## CONCLUSION

Among the most severe diseases, BT affects both young and middle-aged people. Massive amounts of images are generated by the MRI scans. These images are inspected by a radiologist, but due to the complexity of tumours, a human inspection of the MRIs may introduce inaccuracies. Computerised systems for classification powered by artificial intelligence (AI) have frequently outperformed human classification methods. Because of this, we propose a fuzzy logic approach for better BT identification and classification. The study begins with the gathering of data, which is then processed through steps including filtering, binarization, and intensity variation, followed by the extraction of features like size and threshold values, and finally the implementation of a fuzzy-based classification system. The proposed method successfully detects tumours with a 96.0% success rate. In the future, we hope to build a global AI model that can accurately detect tumours in multiple organs, including the breast, liver, and lungs, with minimal error. In conclusion, this research serves as a stepping stone towards more advanced and accurate AI-based medical imaging systems, highlighting the importance of continuing to explore and refine these methodologies. The potential impact of this research on improving patient outcomes and advancing medical diagnostics underscores the significance of investing in further research and development in the field of AI-driven healthcare technologies.

## DISCLAIMER:

None